# RESEARCH ARTICLE

## ARTICLE TITLE: Automatic Transcription of Drum Strokes in Carnatic Music

**Authors:** Kausthubh Chandramouli[a], William Sethares[b]
a: The Hong Kong Polytechnic University
b: University of Wisconsin–Madison

**Abstract**
The *mridangam* is a double-headed percussion instrument that plays a key role in Carnatic music concerts. This paper presents a novel automatic transcription algorithm to classify the strokes played on the *mridangam*. Onset detection is first performed to segment the audio signal into individual strokes, and a feature vectors consisting of the DFT magnitude spectrum of the segmented signal are generated. A multi-layer feedforward neural network is trained using the feature vectors as inputs and the manual transcriptions as targets. Since the *mridangam* is a tonal instrument tuned to a given tonic, tonic invariance is an important feature of the classifier. Tonic invariance is achieved by augmenting the dataset with pitch-shifted copies of the audio. This algorithm consistently yields over 83% accuracy on a held-out test dataset.

Word Count: 3723

## 1. INTRODUCTION

Carnatic music is a system of classical art music associated with the southern states of India, with a long history that can be traced back at least over 500 years. In its present-day form, Carnatic music is primarily presented in the form of concerts performed by small groups consisting of a melodic lead (usually vocals), a melodic instrumental accompaniment (usually a violin), and a rhythmic accompaniment (usually a traditional two-headed percussion instrument called the *mridangam*, often supported by a frame drum called the *kanjira* or a clay percussion instrument called the *ghatam*). A closely related form is *konnakol*, which is a form of verbalization of the percussive patterns played on the different instruments, which is used as a performance form as well as a teaching tool for the *mridangam* or the *kanjira*.

The *mridangam* has a long tradition associated with it and is played using several different strokes on the treble (right) and bass (left) heads. Different finger positions and methods of striking the drumheads produce a variety of tones, with some strokes producing harmonic sounds with a recognizable pitch, and some producing tonic-independent sounds. The number of different strokes varies between different schools of practice, but the basic strokes (that are common across schools) are shown in figure 1. Each stroke produces a unique sound, and this paper presents a way to classify the stroke played based on features extracted from the sound produced.

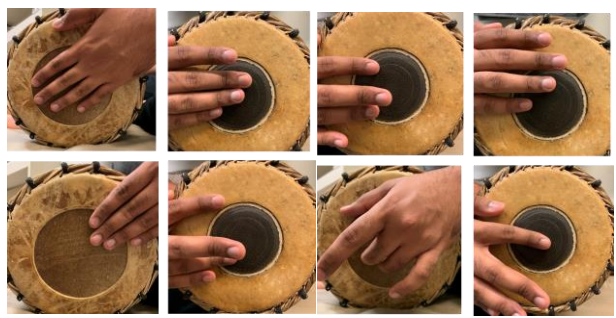

**Figure 1**. Mridangam strokes technique

Our work tackles the problem of automatically transcribing *mridangam* audio recordings into their corresponding strokes and the associated stroke onsets, in a tonic-invariant manner. Automatic transcription is an important step towards generating large, labelled datasets for further research and generating human-readable musical notation from audio. Automatic transcription also serves as the base for 'translation' between different Carnatic percussion instruments, similar to translation between different languages.

## 2. RELATED WORK

Automatic stroke transcription for percussion in Indian classical music is a problem that has been tackled by multiple researchers. Anantapadmanabhan et al. (2013) ini-tially tackled the problem of stroke classification, using a dataset of individual *mridangam* strokes to spectrally analyse the various excitation modes of the *mridangam* when playing different strokes. They then used the modal activations of different strokes to classify them using Hidden Markov Models (HMMs). The disadvantage of this approach is that it requires the modes to be computed for each instrument, and hence is not generalizable to *mridangams* tuned to different tonics, or even to different *mridangams* tuned to the same tonic. However, this issue was tackled in (Anantapadmanabhan et al., 2014), where the authors achieve tonic invariance by using the magnitude spectrum of the constant-Q transform of the signal as input data. They then use non-negative matrix factorization to obtain a low-dimensional feature space where different *mridangam* strokes are easily separable, after which an SVM is used to classify strokes based on those features. Another approach to tackle the problem of tonic-invariant transcription is proposed in (Kuriakose et al., 2015), using HMMs with cent filter cepstral coefficients (CFCCs) as a tonic-invariant input to classify strokes. Kuriakose et al. (2015) also introduced the use of a bigram language model in addition to the HMM to improve transcription performance.

However, the use of constant-Q transforms in (Anantapadmanabhan et al., 2014) and CFCCs in (Anantapadmanabhan et al., 2014) requires careful selection of the range of frequencies to be considered. Additionally, as stated in (Kuriakose et al., 2015), classification performance is relatively poor for tonic-independent, inharmonic strokes since their features are tonic-normalized even though it is not required. As evidenced by the existing body of work, designing a tonic-invariant transcription algorithm is a significant problem, and this work presents a relatively simple, data-driven approach to tonic-invariant transcription with minimal pre-processing and handcrafting.

Neural network-based classifiers for automatic transcription of *tabla* (North Indian drum set) strokes have been explored in (Bhattacharjee et al., n.d.) and (Chordia, 2005), with quite some success. The authors of (Bhattacharjee et al., n.d.) tackled a 4-way classification problem, where the large number of strokes playable on a *tabla* were labelled into 4 distinct categories, using a convolutional neural network (CNN) for classification with a large set of stroke-specific acoustic features as input. They also explored the usage of an ensemble of single-stroke binary classifiers for transcription, achieving f-scores between 70 and 79. A variety of classification models, including feedforward neural networks, multivariate Gaussian distributions, and binary trees were explored in (Chordia, 2005), using multiple datasets with different annotation conventions. A large number of input features that characterize the timbre of the stroke were used, such as temporal centroids, attack time, zero-crossing rate, etc., which were then reduced using principal component analysis, and feedforward neural networks outperformed the other classification models in testing.

The algorithm presented in this work has two components: the first is an onset detection module to extract all the stroke onset times from the given recording, and the second is a neural network classifier to classify strokes based on frequency information around the detected onsets. In the classifier module, the DFT magnitude spectra of the audio signal near the detected onsets are used as a kind of "signature" to identify the stroke. A few other types of features were tested, such as mel-frequency cepstral coefficients (MFCCs), DFT phase spectra, continuous wavelet transform scalograms, etc. However, none of the other features resulted in significantly greater accuracy than the DFT magnitude spectra alone.

## 3. DATASET AND PROCESSING

The dataset we used in this work is a collection of compositions in different rhythmic cycles for Carnatic percussion, written and performed by Akshay Anantapadmanbhan, an independent musician and researcher. The compositions are representative of typical percussion solos in Carnatic music, and were recorded by Mr Anantapadmanabhan on the *mridangam*, *kanjira*, and as *konnakol* in multiple speeds. The dataset consists of 3 compositions in *Adi Tala* (8 beat cycle), 2 compositions in *Misra Chapu* (7 beat cycle), and 1 composition in *Khanda Chapu* (5 beat cycle). Two of the *Adi Tala* compositions are played in 3 different base speeds each (70, 85, and 105 bpm), while all the others are played at a single base speed (70 bpm for *Adi Tala*, 125 bpm for *Misra Chapu*, and 85 bpm for *Khanda Chapu*). In total, there are about 27 minutes of audio in the dataset.

They were then manually annotated with the Sonic Visualizer (Cannam et al., 2010) software, using a reduced set of stroke labels (*lo, hi, mid1-3* for *mridangam* and *lo, mid, hi* for the *kanjira*) to represent the gamut of strokes playable on the instruments. The reduced stroke label set was evaluated to be satisfactorily representative of the wider range of strokes playable on Carnatic percussion instruments by an experienced Carnatic percussionist (Mr Anantapadmanabhan). To classify composite strokes on the *mridangam*, which consist of a stroke played on the treble head along with a stroke played on the bass head (resulting in onset times very close to each other), we generated a $6^{th}$ label (*composite*) for the *mridangam* annotations by binning strokes closer than 0.03s together and labelling them as a single new type of stroke, essentially turning the problem from 5-way to 6-way classification.

The first module in the transcription algorithm is an onset detector. We use a relatively simple onset detection algorithm that uses the location of the peaks of the spectral flux onset strength envelope of the signal to identify stroke onsets (Bello et al., 2005). Spectral flux here refers to the rate of change of the signal power spectrum, and the onset strength envelope is computed as the mean spectral flux across frequencies over time. The peaks in the onset strength envelope correspond to locations in time where the power of the signal changes rapidly, indicating the presence of an onset. An example of the onset strength envelope with the peaks (highlighted in black) is shown in figure 2 below, along with the detected onsets overlaid on the audio signal. A straightforward onset detection method like this can be used because the audio consists of solo *mridangam* only. A more complex signal involving multiple instruments may require more sophisticated onset detection methods such as in (Schluter & Bock, 2014), or in the case of Carnatic percussion specifically, (Manoj Kumar et al., 2015).

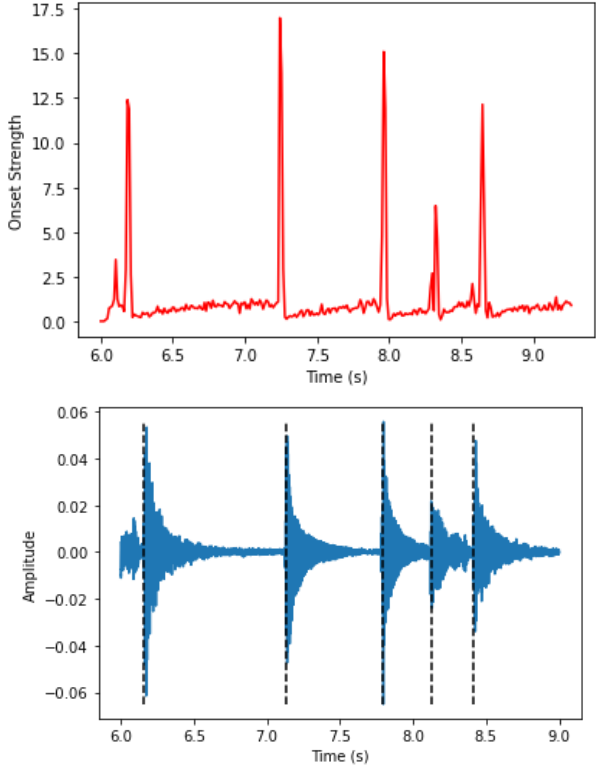

**Figure 2.** Onset detection using onset strength envelopes

To generate the frequency spectra inputs for the neural network, we used a DFT magnitude spectrum computed using a variable-length window near the detected onset times for each stroke. The DFT window extends from 0.03s before the detected onset time of the stroke in question, to 0.03s before the next detected onset time. This allows for maximal frequency resolution, while minimizing spectral leakage between different strokes. The windowed signal is zero-padded up to 48,000 samples to maintain a constant length DFT output as required by the feedforward neural network classifier. We used the magnitude of the DFT spectra up to 12 kHz (with the Nyquist frequency being 24 kHz), resulting in an input vector of length 12,000. An example of the audio signal of a stroke and the corresponding DFT magnitude spectrum is shown below.

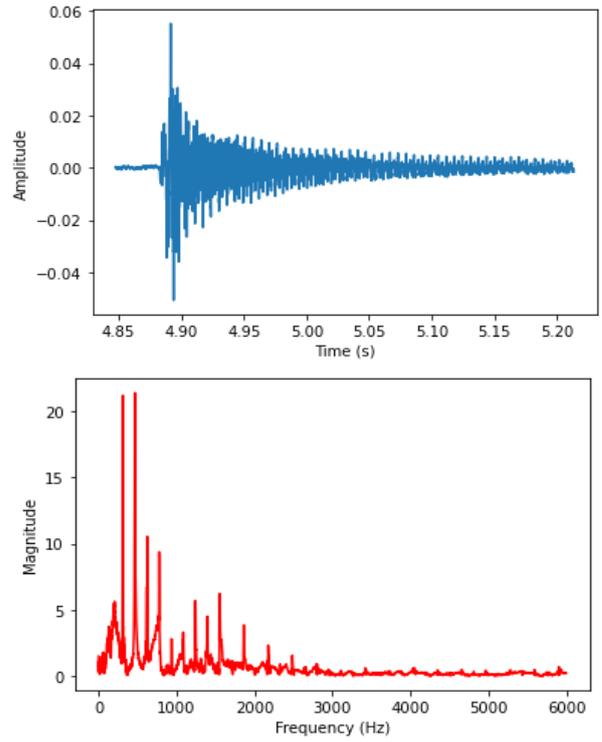

**Figure 3.** Example of a stroke waveform and magnitude spectrum

To train a tonic-invariant classifier, copies of each audio recording pitch-shifted by 1 and 2 semitones upward and downward were used as a form of data augmentation. The same onset detection and spectrum generation algorithms were used, with the same manual annotations used as ground truth values.

## 4. CLASSIFIER

The classifier used in this work is a relatively straightforward multi-layer feedforward neural network. Since frequency spectra are directly used as inputs, a reasonably wide and deep architecture is chosen to maximize expressive ability. The proposed neural network has 7 layers, consisting of an input layer, 5 hidden layers, and an output layer. The input layer has a length of 12,000 (corresponding to the shape of the clipped DFT magnitude spectrum), and the hidden layers have successively lower numbers of neurons as shown in figure 4 below. The output layer, representing class probabilities, has 6 neurons, one for each of the six classes which, when scaled properly can be interpreted as the probability that the given stroke lies in the $n^{th}$ class . All layers used the ReLU (Rectified Linear Unit) activation, with the output layer using a softmax activation

```
Layer (type)                 Output Shape              Param #
=================================================================
input_1 (InputLayer)         [(None, 12000)]           0
dense (Dense)                (None, 15000)             180015000
dropout (Dropout)            (None, 15000)             0
dense_1 (Dense)              (None, 9000)              135009000
dropout_1 (Dropout)          (None, 9000)              0
dense_2 (Dense)              (None, 4500)              40504500
dropout_2 (Dropout)          (None, 4500)              0
dense_3 (Dense)              (None, 1500)              6751500
dropout_3 (Dropout)          (None, 1500)              0
dense_4 (Dense)              (None, 450)               675450
dense_5 (Dense)              (None, 100)               45100
dense_6 (Dense)              (None, 6)                 606
=================================================================
```

**Figure 4.** Classifier architecture summary

Categorical cross-entropy was used as the loss function, as is common for multi-class problems. The model was trained for 25 epochs using the Adam optimizer to implement gradient descent with a learning rate of 0.0002. Early stopping was implemented using the validation accuracy metric to prevent overfitting, and dropout was incorporated to aid generalization. Train and validation sets were generated from the full dataset with an 80/20 split respectively of all the strokes. To further verify the model's performance, the model was also trained on the dataset with a single composition held out, and then tested using the held-out composition (instead of the 80/20 split described above). To test the degree of tonic invariance achieved by the classifier, versions of the classifier were trained on compositions without any data augmentation and tested on pitch shifted data, trained on compositions shifted 1 semitone up and down and tested on data pitch shifted 2 and 3 semitones up and down, as well as trained on compositions shifted 1 and 2 semitones up and down and tested on data shifted 3 semitones up and down.

## 5. RESULTS

The accuracy of the onset detector was assessed based on the percentage of onsets it was able to correctly detect. We define an onset to be correctly detected if the difference between the detected onset time and the annotated "ground truth" value is less than 15 ms. A relatively small threshold was chosen because in typical *mridangam* performance, there are often multiple strokes with very short intervals between them (< 50 ms). Based on this metric, the onset detector used in this work consistently reached accuracies above 90% for all the compositions in the dataset.

he proposed classifier model consistently trains to around 83% accuracy on the 6-way classification problem, with a minimum cross-entropy of around 0.56 on the validation data (see figure 5). As seen in figure X below, the greatest confusion in predictions is between *composite* and *mid3* strokes, as well as between *mid1* and *mid3* strokes, likely caused by the frequent occurrence of *mid3* strokes. The diagonal elements of the confusion matrix also clearly show the class imbalance, with *mid3* and *composite* strokes being much more frequent than the other classes.

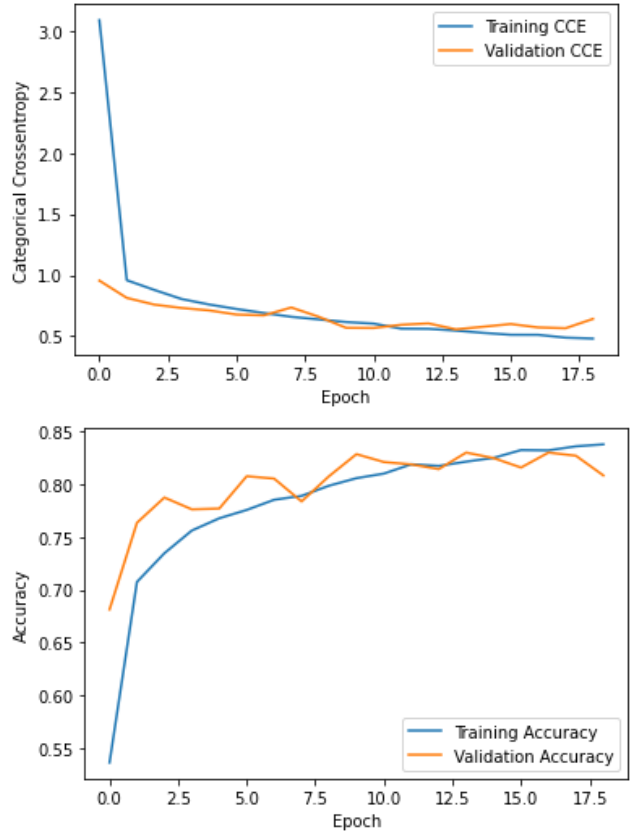

**Figure 5.** Training and validation cross-entropy and accuracy curves

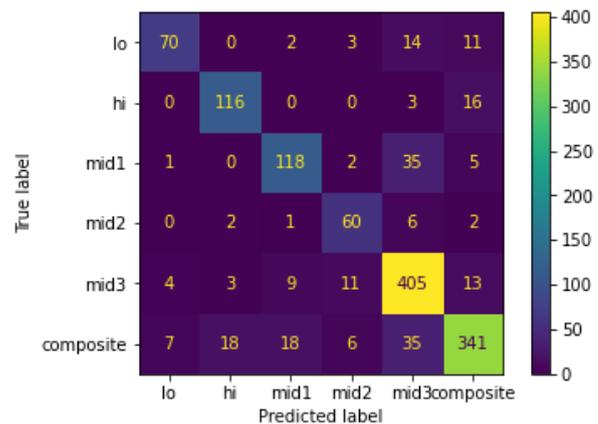

**Figure 6.** Confusion matrix for single-tonic classifier

Table 1 below shows a summary of the results of the tonic-invariance tests:

| Train | Test | Accuracy (%) |
|---|---|---|
| **No** pitch shift data **augmentation** | Previously seen composition shifted **1 semitone** up and down | **71** |
| **No** pitch shift data **augmentation** | Previously seen composition shifted **2 semitones** up and down | **58** |
| Augmentation with **1 semitone** pitch shifts up and down | Previously seen composition shifted **2 semitones** up and down | **68** |
| Augmentation with **1 semitone** pitch shifts up and down | Previously seen composition shifted **3 semitones** up and down | **52** |
| Augmentation with **1 and 2 semitone** pitch shifts up and down | Previously seen composition shifted **3 semitones** up and down | **72** |

**Table 1.** Summary of tonic-invariant classifier training and testing

To explore the effects of the class imbalances in the dataset, we tried building a 'weighted' classifier and a 'balanced' classifier. The weighted classifier was trained with class weights applied to the loss value, such that classes with fewer training examples are weighted more than those with many examples. The class weights were assigned in inverse proportion to the relative frequency of each class. The balanced classifier was trained using a smaller dataset where each class had the same number of training samples. To generate the balanced dataset, 400 examples from each class were selected, since the class with the fewest examples (*mid2*) had just over 400 examples after the train/test split. Both classifiers were tested using the same test dataset used for the baseline classifier, i.e., the test dataset was uniformly sampled from the original stroke dataset, and hence represents the actual relative class frequencies of the full dataset. The confusion matrices (on the test dataset) for the two classifiers are shown below.

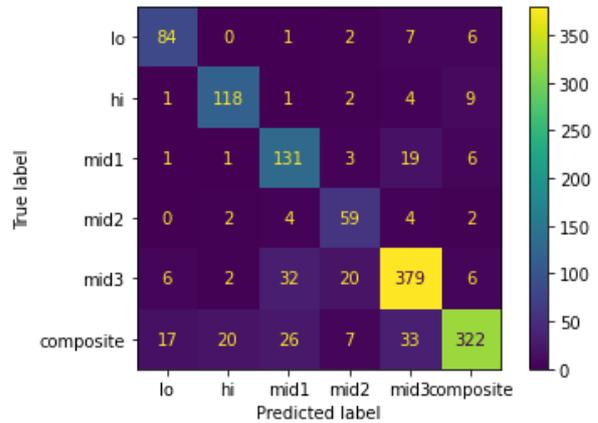

**Figure 7.** Confusion matrix for the 'weighted' classifier

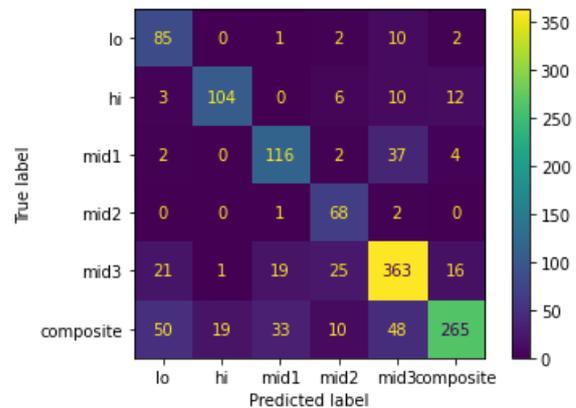

**Figure 8.** Confusion matrix for the 'balanced' classifier

With the same training parameters as the baseline classifier, the weighted classifier trained to 81.7% accuracy, and the balanced classifier to 76% accuracy. Clearly the balanced classifier performs noticeably worse than the baseline and weighted classifiers, but the difference between the weighted and baseline classifiers is more complex. Computing per-class precision and recall scores (shown in tables 2 and 3) from the two confusion matrices, there seems to be a trade-off between precision and recall. The baseline classifier has better recall scores for the most frequent *mid3* and *composite* classes, and better precision scores for the other 4 classes, with the converse holding true for the weighted classifier. The key difference is that the baseline classifier correctly predicts more positive records in the *mid3* and *composite* classes, but since the training data is imbalanced, it also incorrectly labels many strokes as *mid3* or *composite*. Conversely, the weighted classifier has fewer false positives in the *mid3* and *composite* classes. Computing per-class F1-scores for both classifiers, which considers both precision and recall, the differences are small, except for the *mid2* class where the baseline classifier performs significantly better (which may be an outlier since *mid2* has the fewest examples).

| Weighted | *lo* | *hi* | *mid1* | *mid2* | *mid3* | *composite* |
|---|---|---|---|---|---|---|
| *Precision* | 0.77 | 0.83 | 0.67 | 0.63 | 0.85 | 0.92 |
| *Recall* | 0.84 | 0.87 | 0.81 | 0.83 | 0.85 | 0.76 |
| *F1* | 0.80 | 0.85 | 0.74 | 0.72 | 0.85 | 0.83 |

**Table 2.** Classification metrics for 'weighted' classifier

| Baseline | *lo* | *hi* | *mid1* | *mid2* | *mid3* | *composite* |
|---|---|---|---|---|---|---|
| *Precision* | 0.80 | 0.85 | 0.90 | 0.83 | 0.81 | 0.87 |
| *Recall* | 0.79 | 0.84 | 0.65 | 0.77 | 0.94 | 0.84 |
| *F1* | 0.79 | 0.84 | 0.75 | 0.80 | 0.87 | 0.85 |

**Table 3.** Classification metrics for baseline classifier

To compare with a more "traditional" ML classification technique, an SVM classifier was trained using the same single-tonic data, resulting in a validation accuracy of around 74%. For comparison, Anantapadmanabhan et al. (2014) developed a classifier for isolated *mridangam* strokes based on features extracted from constant-Q transform spectra using non-negative matrix factorization, and an overall accuracy of 86.65% is reported. Kuriakose et al. (2015) use features extracted from CFCCs with Hidden Markov Models and a bi-gram language model, and report accuracies of around 88% (correct later). However, both papers use datasets with exact *mridangam* stroke annotations, as opposed to a reduced annotation set. While a reduced annotation set makes the problem simpler, it also means that multiple different physical strokes are mapped to the same label, potentially increasing the complexity of the mapping to be learnt by the classifier.

## 6. DISCUSSION

Based on the accuracy figures for the base classifier trained on data with a single pitch only, we can conclude that the neural network classifier has state-of-the-art performance, while using in a sense "simpler", more low-level inputs than existing algorithms in the literature. When trained with augmented data using pitch shifted audio, the classifier maintains similar accuracy figures of above 80% when tested on held out data and manages to generalize to unseen data from pitches 1 semitone higher or lower as evidenced by the relatively high testing accuracies (>70%) for the same. This suggests that in theory, by using a dataset with recordings made using *mridangams* tuned to different base pitches, a fully general tonic-invariant neural network classifier could be trained. As opposed to existing approaches to tonic-invariant classification in the literature, this method does not require any pre-processing of the frequency spectra to introduce tonic-invariance (which is shown to impact classification performance for inharmonic strokes in (Kuriakose et al., 2015))

To understand the spectral characteristics of each stroke, and better understand the type of information the neural network could be learning from the spectra, we built a set of 'template' spectra (see figure 9) by averaging a large number of spectra from each stroke class. Apart from qualitatively analysing the patterns in the template spectra, we built a simple classifier using the Pearson correlation coefficient between each of the template spectra and a given candidate spectrum as a measure of similarity, selecting the template with the highest correlation as the label for the candidate. This resulted in around 67% classification accuracy, which validates the need for a more sophisticated classifier model while also showing that the template spectra patterns contain a significant amount of information that is likely leveraged during classification. It is interesting to note that the neural network classifier is able to generalize the mapping between the spectra and the labels even when considering training data from multiple different tonics.

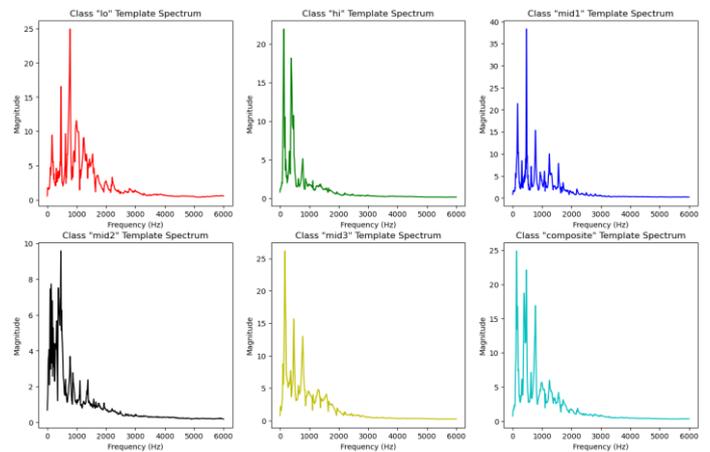

**Figure 9.** Template spectra for each class

From the template spectra, we observe that the spectra often have multiple harmonic peaks - thus verifying/reproducing the observations in Raman's classic paper (Raman, 1935). A large body of literature, continuing from Ramakrishna and Sondhi (1954) to modern explorations such as Sooraj and Padmanabhan (2017) explores the physical origins of the harmonicity and relates the dimensions and materials of the drums and drumheads to the modes of vibration. Missing from such accounts are the various excitation methods (ways of striking the drums) which we explore not by studying the acoustics of the excitation mechanisms (which is likely beyond the present state of the art) but rather in a classification framework where similarities and differences between the drum strokes are "discovered" by the classification algorithms.

One immediate extension of this project would be to train the proposed classifier on a larger dataset that consists of recordings from *mridangams* tuned to different tonics across a larger range (C, C#, D, G, G#, A are among the most commonly used in performance) to verify that the classifier remains tonic invariant even across a wider range of tonics. Additionally, other onset detection algorithms, such as the one proposed in (Manoj Kumar et al., 2015) could be used, which may improve the overall transcription performance. A larger task would be to use the automatic transcription algorithm proposed here as one part of a Carnatic percussion 'translator' that can convert *mridangam* audio into *konnakol* or *kanjira* audio, or vice versa. This would likely require the use of language modelling to capture the complex relationship between the *mridangam* playing form and the spoken *konnakol* form or the *kanjira* playing form.